\input{aipcheck}

\documentclass[
    ,final            
  ]
  {aipproc}

\layoutstyle{6x9}
\begin{document}
\title{Single star-forming galaxies and star-forming galaxies in SF\,+\,SF and mixed pairs}
\author{Birgit Kelm and Paola Focardi}{
  address={Dipartimento di Astronomia, V. Ranzani 1, I-40127 Bologna}
}
\begin{abstract}
We compare the SFR of single star-forming galaxies with the SFR of 
star-forming galaxies in pairs. Volume-limited samples are compared 
selected from the 2dFGRS, applying a maximum magnitude difference criterion. 
We show that SF galaxies in SF\,+\,SF pairs  
typically increase their SFR as they get fainter, whereas this does not 
happen for SF galaxies in mixed (SF\,+\,passive) pairs. 
And we provide evidence that differences between single SF and SF in pairs get 
more significant when SF galaxies in mixed pairs are excluded from the pair 
sample. 
Our analysis confirms that enhanced SFR and the presence of a 
companion galaxy (on 0.5 $h^{-1}$ Mpc scale) are correlated quantities, 
provided the galaxy is neither too luminous nor too faint, and 
the triggering galaxy is itself a SF galaxy.
\end{abstract}
\maketitle
\section{Introduction} 
It is known that the star formation rate (SFR) of galaxies 
decreases as galaxies get more luminous and older, and that 
star formation is inhibited in dense environments such as rich groups and 
clusters (Lewis et al. 2002, Gomez et al. 2003). 
In contrast, galaxy-galaxy interactions seem to be powerful 
mechanisms to trigger the SFR. 
It is clear that interactions occur, however the frequency of these events, 
and the distribution of galaxy luminosities (and masses) involved has yet to 
be established.  
Isolating the relevant quantity that distinguishes between 
single and pair systems is not straightforward. Distance to the companion is 
argued as being the fundamental factor, 
and an extremely close companion has been shown to enhance the SFR of 
galaxies significantly (Lambas et al. 2003, Barton et al. 2003, Nikolic et al. 2004). 
However, such close companions are rare in the present universe, meaning that 
this kind of interaction is not a relevant phenomenon for the galaxy 
population as a whole (Bergvall et al. 2003). In this study we allow 
a distance as large as 0.5 $h^{-1}$ Mpc between pair members. 
\section{Sample selection criteria}
Each 2dF galaxy spectrum is typed on the basis of the relative strength 
of its first two principal components, which are the emission and the absorption components within the spectrum. The parameter $\eta$ is the linear combination of these two components (Madgwick et al. 2002). 
Low $\eta$ are typically early type galaxies, high $\eta$ late type galaxies. 
Qualitatively $\eta$ is an indicator of the ratio of the present to the past 
star formation activity of each galaxy. 
To separate passive from star-forming galaxies we divide our sample at 
$\eta$ = -1.4, as in Madgwick et al. (2003):
 
\begin{description}

\item[$\eta$  $<$ -1.4] ==> Passive

\item[$\eta$  $\geq$ -1.4] ==> Star-Forming 

\end{description}
 
The single SF and the pair SF galaxy samples have been selected from 
the 2dFGRS applying the following criteria:

\begin{description}

\item[pair]  $-->$ 1 companion galaxy within 500$h^{-1}$kpc projected distance 
and $\pm$ 1000km/s depth. No further companion within [0.5 - 1.0]$h^{-1}$Mpc 
projected distance and $\pm$ 1000 km/s depth.  

\item[single]  $-->$ no companion galaxy within 500$h^{-1}$kpc projected 
distance and $\pm$ 1000km/s depth. No further companion within  
[0.5 - 1.0]$h^{-1}$Mpc projected distance and $\pm$ 1000 km/s depth
\end{description}

The large search radius used to find companions 
allows us to select samples of single and pair galaxies that are 
large. 
The request for no companions in the [0.5 - 1]$h^{-1}$Mpc region, for both 
samples, reduces contamination by single and pair galaxies located 
in groups and avoids misinterpretation between the triggering 
effect of one single companion and the triggering effect of multiple 
interactions. It has been shown (Alonso et al. 2004) that pairs embedded 
within groups have a lower SF activity than other group members. 

Companions are here counted applying a maximum magnitude criterion: 
all galaxies that are in a 2-magnitude bin, ranging from the 
absolute magnitude of the galaxy itself to 2 magnitudes fainter, are 
counted as companions. 
To cover the same depth in absolute magnitude space around each SF galaxy,   
only galaxies enter the samples that would still be in the 2dF 
if they were 2 magnitudes fainter. 

To avoid a selection bias, comparison between single SF and SF galaxies in 
pairs is performed using 10 different volume limited samples, 
covering a 1-Magnitude range each. 
The $\eta$ parameter is used to compute k-corrections 
(Madgwick et al. 2002), that are then applied to each galaxy individually to 
derive its absolute magnitude. A flat Lambda cosmology is assumed 
throughout with $\Omega$$_M$=0.3, $\Omega$$_{\Lambda}$=0.7 and h = H/75 km s$^{-1}$ Mpc$^{-1}$.     
\section{Single SF galaxies and SF galaxies in pairs}
In Table 1 we list the number of SF galaxies and the median, upper quartile 
and lower quartile of the $\eta$ distributions, for single and pair samples. 
The fraction of SF galaxies in pairs (normalized to the sum of SF galaxies 
in single and pair systems) is $\sim$30\% in all, but the faintest and 
brightest volume limited samples, where it decreases to 20\%. 

In Figure 1 we plot $\eta$ as a function of absolute 
magnitude, for single SF galaxies (squares) and for SF galaxies in pairs 
(circles).
It clearly emerges that differences between single and 
pair samples are definitely modest when compared to differences 
that both samples undergo because of the dramatic effect of luminosity 
on the SFR of galaxies. 
Nevertheless, Figure 1 indicates that 
the median SFR of SF galaxies in pairs, is either similar or above the median 
values for single SF galaxies. 

We use the KS test to eveluate how significant differences are 
between  single SF galaxies and SF galaxies in 
pairs. We perform the comparison in each volume-limited sample separately. 
The null hypothesis that the single SF and the SF in pair samples are drawn 
from the same parent distribution can be rejected with a significance level 
that is listed in column 2 in Table 3.  

The KS test indicates that, in general, we cannot reject the hypothesis 
that single SF galaxies and SF galaxies in pairs are drawn from the same 
parent population.  
The 2 $\sigma$ significance level is reached in 2 samples only, namely the 
[--19.0 --20.0] and the [--19.5 --20.0] ones.
\begin{figure}
  \includegraphics[height=.4\textheight]{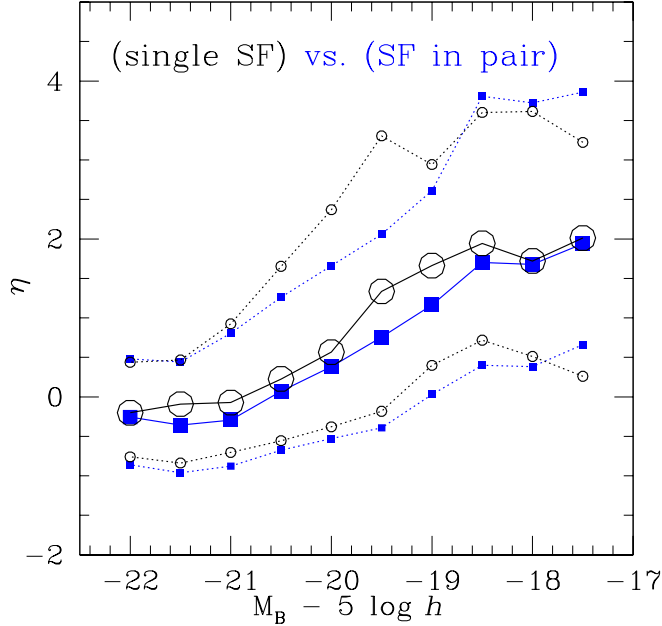}
  \caption{Single SF galaxies (filled squares) and SF galaxies in pairs (circles). 
Median values, upper and lower quartiles are plotted for 10 volume-limited 
samples.}
\end{figure}
\section{SF galaxies in SF\,+\,SF and mixed pairs}
To check whether the triggering effect of a companion galaxy is linked 
to the spectral-type of the companion itself, we next compare  
SF galaxies whose companion is another SF galaxy with SF galaxies 
whose companion is a passive galaxy. 

In Table 2 the number of SF galaxies in SF\,+\,SF pairs 
and the number of SF galaxies in SF\,+\,passive pairs is listed, for each 
volume-limited sample. The incidence of mixed pairs is generally small in 
intermediate luminous samples, and definitely negligible in faint 
(M - 5log{\it h} $>$--19) samples. In general we expect the incidence of 
mixed pairs to be low, due to our choice to select 'isolated' single and pair 
systems (no companion in the 0.5-1.0 h$^{-1}$Mpc region). 

Figure 2 shows median values only for samples that include a minimum 
of 10 sources. Figure 2 shows that 
 SF galaxies in SF\,+\,SF pairs (empty triangles)   
typically increase their SFR as they get fainter, whereas this does not 
happen for SF galaxies in mixed (SF\,+\,passive) pairs (filled triangles). 
Figure 2 thus provides evidence that a passive companion does not trigger 
star formation as efficiently as a SF companion. 
However, the spectral-type of the companion appears to 
relate to the average SFR in intermediate luminous samples only, 
not in the more luminous ones.  
Our results can be compared with those of Nikolic et al. (2004). 
They have investigate the effect of galaxy interaction on 
star formation using the SDSS, and find no dependence 
of star formation enhancement on the morphological type or mass 
of the companion galaxy. 
  
\begin{figure}
  \includegraphics[height=.4\textheight]{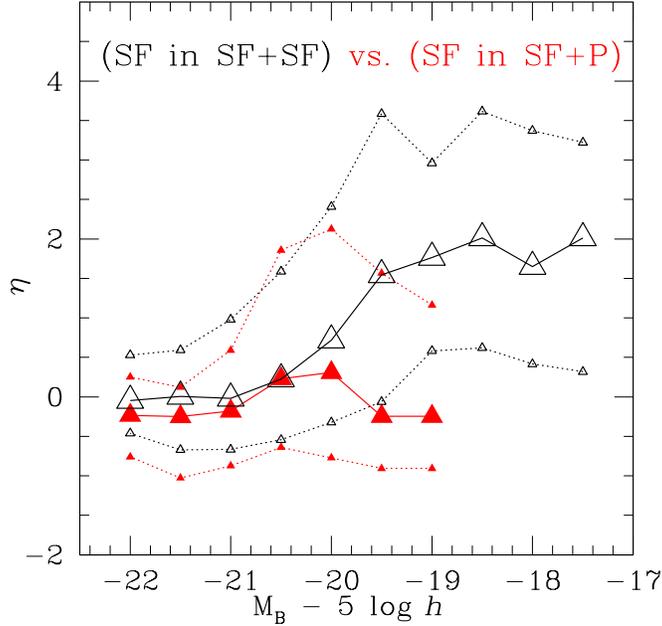}
  \caption{SF galaxies in SF+SF pairs (open triangles) and SF galaxies in SF + Passive pairs (filled triangles). Median values, upper and lower quartiles are plotted for 10 volume-limited 
samples.} 
\end{figure}

\begin{figure}
  \includegraphics[height=.4\textheight]{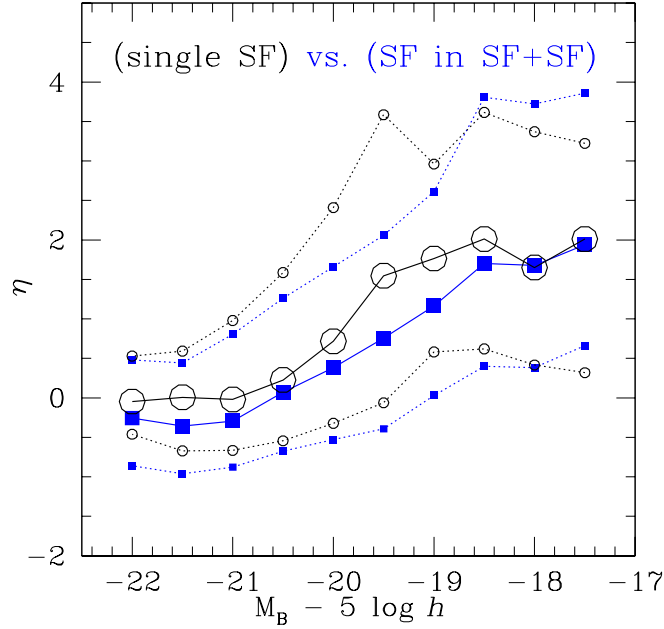}
    \caption{Single SF galaxies (filled squares) and SF galaxies in SF\,+\,SF pairs 
(circles). Median values, upper and lower quartiles are plotted for 10 
volume-limited samples. }
\end{figure}
\section {Single SF galaxies and SF galaxies in SF\,+\,SF pairs}
Having shown that the triggering effect of passive and SF companions 
is different, we next investigate whether differences between 
single SF galaxies and SF galaxies in pairs get more significant 
when SF galaxies in mixed pairs are excluded from the pair sample.   
Figure 3 is the equivalent of Fig. 1 for SF galaxies in SF\,+\,SF pairs 
only. Comparison with Fig.1 reveals that differences between median values 
get more pronounced. 

Again we apply the KS test to evaluate how significant differences 
are. The null hypothesis that the single SF and the SF in the SF\,+\,SF 
samples are drawn from the same parent population can be rejected with a 
significance level that is listed in column 3 in Table 3.  
Differences are significant at 3$\sigma$ c.l. in the [--19.0 --20.0] sample, 
and at the 2$\sigma$ c.l. in the [--18.5 --19.5], [--19.5 --20.5], [--20.5 
--21.5] and [--21.0 --22.0] samples. Differences between 
single SF and pair SF galaxies are not significant, however, when faint  
or extremely bright samples are compared. 

Comparing single SF with SF\,+\,SF pairs improves differences between 
single and pair samples in as many as 4 cases (out of 6 among those having 
L$_{KS}$$\geq$1). We conclude that, when excluding SF galaxies in mixed pairs, 
the SFR of SF galaxies in pairs is larger than the SFR of the single SF 
galaxies.   
\begin{table}
\begin{tabular}{lrcrc}
\hline
   \tablehead{1}{r}{b}{Magnitude\\range}
  & \tablehead{1}{r}{b}{Single SF\\ }
  & \tablehead{1}{c}{b}{$\eta$\tablenote{median - lower quartile - upper quartile} \\}
  & \tablehead{1}{r}{b}{SF in pair \\ }
  & \tablehead{1}{c}{b}{$\eta$ \\}\\
\hline
--17.0 --18.0 &  71 &  1.939 / 0.662  / 3.858 &  18 & 2.012 / 0.260 / 3.223 \\
--17.5 --18.5 & 113 &  1.677 / 0.384  / 3.721 &  48 & 1.720 / 0.511 / 3.614 \\
--18.0 --19.0 & 194 &  1.701 / 0.401  / 3.807 &  97 & 1.944 / 0.718 / 3.602 \\
--18.5 --19.5 & 300 &  1.165 / 0.030  / 2.611 & 128 & 1.663 / 0.399 / 2.942 \\
--19.0 --20.0 & 433 &  0.758 / -0.393 / 2.058 & 160 & 1.336 / -0.183 / 3.305 \\
--19.5 --20.5 & 635 &  0.381 / -0.527 / 1.658 & 234 & 0.566 / -0.392 / 2.369 \\
--20.0 --21.0 & 605 &  0.071 / -0.674 / 1.262 & 252 & 0.227 / -0.554 / 1.653 \\
--20.5 --21.5 & 428 & -0.293 / -0.878 / 0.805 & 176 & -0.072 / -0.704 / 0.925\\
--21.0 --22.0 & 253 & -0.358 / -0.961 / 0.443 &  98 & -0.092 / -0.837/ 0.466\\
--21.5 --22.5 & 113 & -0.256 / -0.860 / 0.481 &  32 & -0.201 / -0.759 / 0.439\\
\hline
\end{tabular}
\caption{Single SF galaxies and SF galaxies in pairs: sample size and average SFR}
\label{tab:a}
\end{table}

\begin{table}
\begin{tabular}{lrcrc}
\hline
   \tablehead{1}{r}{b}{Magnitude\\range}
  & \tablehead{1}{r}{b}{SF\,+\,SF \\}
  & \tablehead{1}{c}{b}{$\eta$ \\}
  & \tablehead{1}{r}{b}{SF\,+\,Passive \\}
  & \tablehead{1}{c}{b}{$\eta$\\}\\
\hline
--17.0 --18.0 &  17 & 2.012 / 0.318 / 3.222 &   1  & - - - \\
--17.5 --18.5 &  46 & 1.651 / 0.416 / 3.371 &   2  & - - - \\
--18.0 --19.0 &  94 & 2.012 / 0.620 / 3.614 &   3  & - - - \\
--18.5 --19.5 & 118 & 1.764 / 0.581 / 2.960 &   10  & -0.246/ -0.906 / 1.164\\
--19.0 --20.0 & 135 & 1.546 / -0.062 / 3.586 &  25  & -0.247/ -0.906 / 1.566\\
--19.5 --20.5 & 188 & 0.717 / -0.322 / 2.410 &  46  & 0.309 / -0.772 / 2.125\\
--20.0 --21.0 & 188 & 0.227 / -0.543 / 1.587 &  64  & 0.230 / -0.640 / 1.858\\
--20.5 --21.5 & 132 & -0.020 / -0.665 / 0.979 & 44  & -0.180 / -0.873 / 0.591\\
--21.0 --22.0 &  65 & 0.007 / -0.672 / 0.593 &  33 & -0.250 / -1.027 / 0.120\\
--21.5 --22.5 &  20 & -0.046 / -0.460 / 0.529 & 12 & -0.234 / -0.759 / 0.251\\
\hline
\end{tabular}
\caption{SF galaxies in SF\,+\,SF and mixed pairs: sample size and average SFR}
\label{tab:a}
\end{table}

\begin{table}
\begin{tabular}{lrr}
\hline
   \tablehead{1}{r}{b}{Magnitude\\range}
  & \tablehead{1}{r}{b}{SF single vs. SF in pair\\L$_{KS}$}
  & \tablehead{1}{r}{b}{SF single vs. SF + SF \\ L$_{KS}$}\\
\hline
--17.0 --18.0 &  $<$1      & $<1$ \\
--17.5 --18.5 &  $<1$      & $<1$ \\
--18.0 --19.0 &  $<1$      & $<1$ \\
--18.5 --19.5 &   1        &  {\bf 2} \\
--19.0 --20.0 &  {\bf 2}  &  {\bf 3}  \\ 
--19.5 --20.5 &  {\bf 2}  &  {\bf 2}  \\
--20.0 --21.0 &   1        &   1 \\
--20.5 --21.5 &   1        &   {\bf 2} \\
--21.0 --22.0 &   1        &   {\bf 2} \\
--21.5 --22.5 &   $<$1     &   $<$1 \\
\hline
\end{tabular}
\caption{KS-test: the confidence level (in units of standard deviation 
$\sigma$) is shown with which we reject the null hypothesis that the single SF and the SF in pair 
samples are drawn from the same parent distribution. 
The value in col. 2 refers to SF galaxies in all pairs, the value in col. 3 
to SF galaxies in SF\,+\,SF pairs only}
\label{tab:a}
\end{table}
\bibliographystyle{aipproc}   
\section {References}

{\hskip 0.32truecm} Alonso M.S., Tissera P., Coldweel G. \& Lambas D.G. 2004, MNRAS, 352, 1081

Barton E., Geller M.J. \& Kenyon S.J. 2003, ApJ, 582, 668

Bergvall N., Laurikainen E. \& Aalto S. 2003, A\&A, 405, 31 
 
Gomez P.L., Nichol R.C., Miller C.J et al. 2003, ApJ, 584, 210 

Lambas D.G., Tissera P.B., Alonso M.S., Coldwell G. 2003, MNRAS, 346, 1189

Lewis I., Balogh M., De Propris R. et al. 2002, MNRAS, 334, 673

Madgwick D.S., Lahav O., Baldry I.K. et al. 2002, MNRAS, 333, 133

Madgwick D.S., Hawkins E., Lahav O. et al. 2003, MNRAS, 344, 847

Nikolic B., Cullen H. \& Alexander P. 2004, [astro-ph/0407289]
\IfFileExists{\jobname.bbl}{}
 {\typeout{}
  \typeout{******************************************}
  \typeout{** Please run "bibtex \jobname" to optain}
  \typeout{** the bibliography and then re-run LaTeX}
  \typeout{** twice to fix the references!}
  \typeout{******************************************}
  \typeout{}
 }

\end{document}